\documentstyle[PASJadd]{PASJ95}
\input psfig

\begin{document}
%%%%%%%%%%%%%%%%%%%%%%%%%%%%%%%%%%%%%%%%%%%%%%%%%%%%%%%%%%%%%%%%%%%%%%%%%%%
%\special{!userdict begin /bop-hook{gsave 150 90 translate
%0 rotate /Times-Roman findfont 50 scalefont setfont
%-145 -70 moveto 0.9 setgray
%(DRAFT - DRAFT - DRAFT) show
%grestore}def end}
%%%%%%%%%%%%%%%%%%%%%%%%%%%%%%%%%%%%%%%%%%%%%%%%%%%%%%%%%%%%%%%%%%%%%%%%%%%%
\setcounter{page}{1}

\title{Accretion Disks Phase Transitions: 2-D or not 2-D?}

\author{Marek A. \ {\sc Abramowicz}\thanks{%
Nordita, Blegdamsvej 17, DK-2100 Copenhagen \O, Denmark.}~\thanks{% 
Laboratorio Interdisciplinare SISSA, Trieste, Italy, and ICTP, Trieste, Italy.
} \\
{\it Institute of theoretical physics, G{\"o}teborg
           University and Chalmers University of Technology,
           412 96 G{\"o}teborg, Sweden} \\
{\it E-mail(MAA): marek@fy.chalmers.se}  % colon was missing
\\[6pt]
Gunnlaugur {\sc Bj\"ornsson} \\
{\it Science Institute, Dunhagi 3, University of Iceland,
           IS-107 Reykjavik, Iceland} \\%[6pt]
%C.\ P.\ {\sc Dullemond} \\
%{\it Leiden Observatory, P. O. Box 9513, 2300 RA Leiden, The Netherlands} \\
and \\
Igor V.\ {\sc Igumenshchev} \\
{\it Institute of Astronomy, 48 Pyatnitskaya Street, Moscow, 109017, Russia}}

\abst{We argue that the proper way to treat thin--thick
accretion-disk transitions
should take into account the 2-D nature of the problem.
We illustrate the physical inconsistency of the 1-D vertically integrated
approach by discussing a
particular example of the convective transport of energy.}

\kword{accretion, accretion disks --- convection --- hydrodynamics}

\maketitle
\thispagestyle{headings}

%\section
%{Introduction}

Traditionally, accretion disks are modeled using a dimensional splitting
method, in which the vertical structure is solved locally at each
radius. Although this procedure is evidently correct for thin disks, it
may be questionable for thick disks 
that have a non-negligible geometric thickness, and for thin or thick
disks experiencing a sharp transition (stationary or non-stationary)
from one state to another.

Advection-dominated accretion flows (ADAFs) are now widely regarded as
being the source of Comptonized X-ray radiation observed from many X-ray
binary sources. It is believed that in such a source, the ADAF
is located in the inner part of an accretion disk around a compact object,
while the outer part is most likely a standard Shakura--Sunyaev disk
(SSD). A transition from the SSD to an ADAF may occur at hundreds or
thousands of gravitational radii from the central compact body. The
physical cause of the transition remains elusive.
1-D ADAF models based on vertical integration are
remarkably successful in explaining the observed spectra
(see e.g. Narayan et al. 1996; Narayan et al. 1997a).
It is easy to overlook, however, that this by itself does not
prove that these models are able to correctly describe the SSD--ADAF
transitions.

Indeed, in this Note we argue that 1-D methods that have been used to
study ADAFs, cannot describe transitions between ADAFs and SSDs. The
problem here is not only with accuracy, but also with the very physical
consistency of these methods. For this reason, several recent ideas and
results concerning the SSD--ADAF transitions cannot be trusted.

We illustrate the physical inconsistency of the vertical integration
approach, which is the most often used 1-D method, by discussing a
particular example of the convective transport of energy. In principle,
one may argue (as e.g. Honma 1996a) that a convective energy flux that
originates in the inner part of the ADAF  could be deposited in inner edge
of the outer SSD, causing its `evaporation' thereby sustaining the
SSD--ADAF transition. In reality, most of the convective flux goes in the
vertical direction and never reaches the SSD. By vertical integration one
artificially forces all of this flux to be deposited in the SSD.

The physical inconsistency of the 1-D methods should not be confused with
the fact that the global models constructed by such methods could be
formally self-consistent.
Indeed, the formal self-consistency of the 1-D
ADAF models
is connected to the fact that the vertically integrated method always
produces a smooth radial dependence of the scale height in global
solutions (Narayan et al. 1997b; Chen et al. 1997).
However, it is most
likely that SSD--ADAF transitions are far from being smooth, as has been
demonstrated in 2-D hydrodynamics simulations by Abramowicz, Igumenshchev,
and Lasota (1998). In this paper, we extend the argument by pointing out
that the two-dimensional nature of the transition is also important for
the heat flux. We consider a particular case of a convective heat
flux, because it is known to be important for ADAFs (Narayan, Yi 1994;
Igumenshchev et al. 1996).

% Fig.1
\begin{figure}
\hbox{\psfig{file=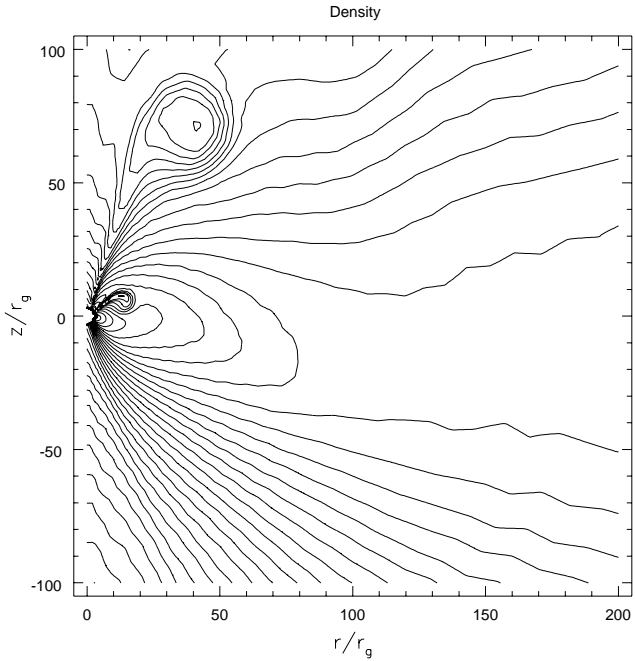,height=8.7cm}}
\caption
%\begin{fv}{1}{18pc}%
{Density distribution of ADAF (viscosity parameter $\alpha=0.1$)
in the meridional
cross-section. The vertical axis coincides with the axis of rotation.
The black hole is located at the origin ($0,0$).
Axes are labeled in the units of $r_g$.
The contours of density $\rho$ are spaced with
$\Delta\log\rho=0.1$.
The time-dependent calculation starts from an initial state,
and after a few characteristic accretion times the quasi-periodic
behaviour of the flow is reached.
The accretion flow is convectively unstable and, as a result,
hot convective bubbles quasi-periodically originate in the innermost
region of the flow.
Two subsequent convective bubbles are clearly seen in the upper hemisphere
of the model.
The bubbles consist of hot and low density matter, and move outwards
under action of the buoyancy force, deviating in the polar direction.
Typically, the subsequent convective bubbles
move outwards in different directions, in the upper and lower hemispheres.}
%\end{fv}
\end{figure}

In order to evaporate the SSD at a certain radius and sustain an SSD--ADAF
transition, a significant amount of energy must be pumped into the
region where the evaporation takes place. The physical mechanism
for that is still unclear, and several possibilities have been suggested,
among them radial convective heat flux. It is precisely here that the
1-D vertically integrated methods fail dramatically.
A treatment based
on vertical integration artificially channels all of the convective flux
to the inner edge of the SSD. This results in a gross over estimation of
the amount of heat pumped into the SSD, and would allow the disk to
evaporate, even when in reality this is not possible. The reason is simply
because convective flux is carried by convective bubbles, which have
the tendency to flow over the disk, rather than converge into the disc
inner rim and release their energy there. Moreover, as the bubbles move
towards larger radii, their motion deviates from the equatorial plane.
Some of the bubbles may even leave the ADAF interior, escaping in roughly
polar directions. Figure~1 presents an example snap-shot of such a behaviour of
the convective bubbles that was calculated with a help of a fully 2-D,
time dependent, hydrodynamical simulation. The bubbles quasi-periodically
originate due to convective
instability; for detail see Igumenshchev and Abramowicz (1999). Because the
bubbles have a tendency to miss the SSD inner rim, they cannot insert
any appreciable amount of energy into the evaporation region.

ADAF models cannot extend to infinity because they are systems with
a negative total energy. Thus, at a certain radius, the disk thickness
drops sharply. This behaviour is similar to geometrically thick
tori, which have a negative total energy.
2-D simulations of viscous evolution of thick tori given by
Igumenshchev et al. (1996) and Stone, Pringle, and Begelman (1999)
have shown that the convective heat flux is not focused toward the
equatorial plane near to the outer boundary of tori where the disk
thickness drops sharply. Thus, focusing of the convective
heat flux into the equatorial region is a consequence
of the 1-D treatment.
A schematic illustration of the artificial focusing effect 
in 1-D approach is presented in figure~2.

% Fig.2
\begin{figure}
\hbox{\psfig{file=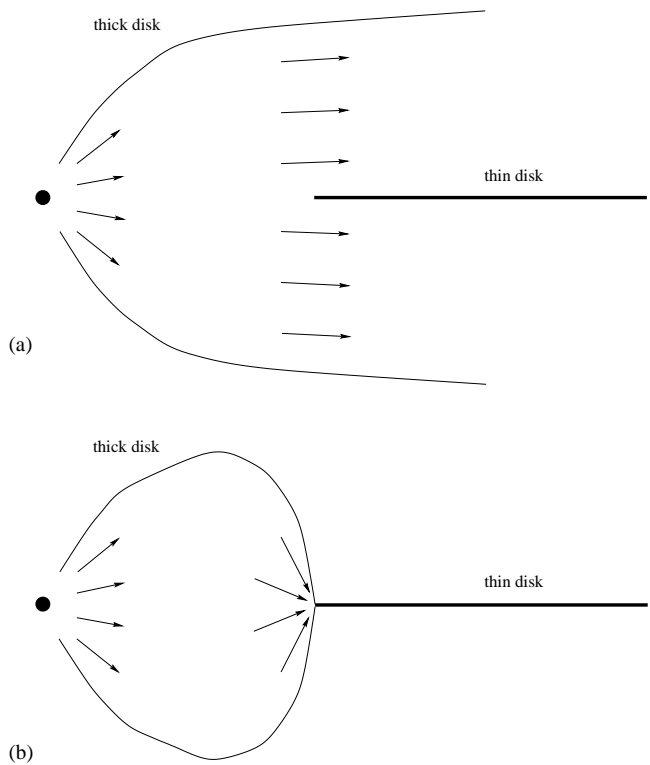,width=8.5cm}}
\caption
%\begin{fv}{2}{18pc}%
{(a) The main part of outward directed energy flux (shown by arrows)
misses the inner rim of thin accretion disk.
(b) Artificial focusing of the energy flux in a 1-D vertically integrated
approach.}
%\end{fv}
\end{figure}

We admit that the present paper does not directly prove that the
behaviour assumed in the Honma model cannot take place, but it is
important to point out its potentially serious difficulty. The Honma model
is now one of the most clever and important ideas in accretion-disk
theory. For this reason, the criticism of it that we provide
should be known to the community.

A similar problem arises when the radial energy flux is provided by radiation.
If vertically integrated intensities (or moments of intensity) are used
to model the flux, the heat input into the SSD is overestimated,
since in such models the inner rim of the SSD by construction
absorbs all of the radiation. In reality, most of the radiation flux passes
the SSD.

The two-dimensional nature of the SSD--ADAF transition is overlooked in
an approach based on vertical integration. Indeed, in a number of
papers on global solutions of vertically integrated accretion disk,
models of sharp transitions between an SSD and an ADAF have been
presented, without sufficiently mentioning the limitations of the
1-D approach (e.g. Narayan, Popham 1992; Honma 1996a, 1996b;
Manmoto et al. 2000).

The vertically integrated 1-D
calculations of thermal fronts propagation in thin disks
meet a similar problem (for the models of cataclysmic variables
see Hameury et al. 1998, and references therein; for
the limit-cycle models see Honma, Matsumoto, and Kato 1991;
Szuszkiewicz, Miller 1997, 1998).
In these models the thermal instability
causes the formation of a geometrically thick and hot inner region of the
accretion disk, which
expands outward, forcing (artificially, due to assumptions)
the outer, Shakura--Sunyaev type, region of the disk
to change its state in the narrow transition region.
Although it could be that the 2-D consideration
of the transition region in these cases
will not change the results of 1-D approach drastically,
because of a small contrast (a factor of few) of the thicknesses of perturbed
and unperturbed parts of the disk, only 2-D models could eventually
remove all doubts about the physical validity of the methods that
are now commonly used.

Another example of the problem of disk transition solved in the 1-D approach
was given by Meyer and Meyer-Hofmeister (1994), who
use the thermal conduction in the vertical direction
to transport energy
into the evaporation region.
Again, this reduces the full 2-D problem to 1-D and excludes
from consideration
the very important factor of the radial energy conductive flux.

We argue that the proper way to treat the
thin--thick disk transitions, in all astrophysical contexts mentioned above,
should take into account the 2-D nature of the problem, and that the 1-D
methods that have been used so far are not satisfactory in this respect.
Unless someone proposes a clever new method to incorporate the relevant
physics into the vertically integrated equations, the only
remaining alternative is to use 2-D numerical simulations.

\par
\vspace{1pc}\par
We thank an anonymous referee for strong criticism which helped us to improve
the paper.
GB acknowledges partial support from the University of Iceland Research Fund.

\section*{References}
\small

\re
Abramowicz M. A., Igumenshchev I. V., Lasota J.-P. 1998, MNRAS 293, 443
\re
Chen X., Abramowicz M. A., Lasota J. P. 1997, ApJ 476, 61
\re
Hameury J.-M., Menou K., Dubus G., Lasota J.-P., Hure J.-M. 1998, 
MNRAS 298, 1048
\re
Honma F. 1996a, in Physics of accretion disks, 
ed S. Kato, S. Inagaki, S. Mineshige, J. Fukue (OPA, Amsterdam B.V.) p 31
\re
Honma F. 1996b, PASJ 48, 77
\re
Honma F., Matsumoto R., Kato S. 1991, PASJ 43, 147
\re
Igumenshchev I. V., Abramowicz M. A. 1999, MNRAS 303, 309
\re
Igumenshchev I. V., Chen X., Abramowicz M. A. 1996, MNRAS 278, 236
\re
Manmoto T., Kato S., Nakamura K. E., Narayan R. 2000, ApJ 529, 127
\re
Meyer F., Meyer-Hofmeister E. 1994, A\&A 288, 175
\re
Narayan R., Barret D., McClintock J. E. 1997a, ApJ 482, 448
\re
Narayan R., Kato S., Honma F. 1997b, ApJ 476, 49
\re
Narayan R., McClintock J. E., Yi I. 1996, ApJ 457, 821
\re
Narayan R., Popham R. 1993, Nature 362, 820
\re
Narayan R., Yi I. 1994, ApJ 428, L13
\re
Stone J. M., Pringle J. E., Begelman M. C. 1999, MNRAS 310, 1002
\re
Szuszkiewicz E., Miller J. C. 1997, MNRAS 287, 165
\re
Szuszkiewicz E., Miller J. C. 1998, MNRAS 298, 888

\label{last}
\end{document}